\documentclass[preprint]{ptephy_v1}

\usepackage{url}

\begin{document}

\title{ Evaluation of radon adsorption
efficiency values in xenon with activated carbon fibers} 

\author[1]{Y.~Nakano}
\author[2,3]{K.~Ichimura}
\author[4]{H.~Ito}
\author[4]{T.~Okada}
\author[4,3]{H.~Sekiya}
\author[1,3,*]{Y.~Takeuchi}
\author[4]{S.~Tasaka}
\author[4,3]{M.~Yamashita}

\affil[1]{Department of Physics, Graduate School of Science, Kobe University, Kobe, Hyogo 657-8501, Japan}
\affil[2]{Research Center for Neutrino Science, Tohoku University,
Sendai, Miyagi 980-8578, Japan}
\affil[3]{Kavli Institute for the Physics and Mathematics of the
Universe (WPI), The University of Tokyo Institutes for Advanced Study,
University of Tokyo, Kashiwa, Chiba 277-8583, Japan} 
\affil[4]{Kamioka Observatory, Institute for Cosmic Ray Research, The
University of Tokyo, Gifu 506-1205, Japan 
\email{takeuchi@phys.sci.kobe-u.ac.jp}}

\begin{abstract}
 The radioactive noble gas radon-222~($\mathrm{^{222}Rn}$, or Rn) produced in the uranium series is
 a crucial background source in many underground experiments.
 We have estimated the adsorption property of Rn with Activated Carbon
 Fibers~(ACFs) in air~(Air), argon~(Ar), and xenon~(Xe) gas.
 In this study, we evaluated six ACFs, named A-7, A-10, A-15, A-20, A-25,
 and S-25, provided from UNITIKA Ltd.
 We measured intrinsic radioactivity of these ACF samples, and found
 A-20's radioactivity of the uranium series
 is $<5.5$~$\mathrm{mBq/kg}$ with $90\%$ confidence level.
 In Air and Ar gas, we found  ACF A-15 has the adsorption efficiency of
 $1/10000$ reduction at maximum before saturation of Rn adsorption,
 and more than $97\%$ adsorption
 efficiency after the saturation. 
 In Xe gas, we found ACF A-20 has the best Rn adsorption ability
 among tested ACFs.
We also found S-25, A-25, and A-15 have similar Rn adsorption performance.

 \end{abstract}

\subjectindex{H20, C43}

\maketitle

\section{Introduction} \label{sec_intro}

\subsection{Xenon for underground experiment and radon background}

Xenon~(Xe) is one of the attractive materials in the field of a particle
physics experiment. Since Xe has no long-lived radioactive isotopes of
its own, Xe is intrinsically radio-pure, except for the double beta
decay ($^{136}$Xe) and double electron capture ($^{124}$Xe) nuclide~\cite{kamlandzen, xenondec}.
Because of its nuclear properties~(such as spin, atomic mass number) as well as its ease of scalability,
 Xe is widely used as a target of
direct dark matter searches~\cite{lux, xenon100, xmass, panda} and a
source of double beta decay searches~\cite{kamland, exo, next, axcel} in underground experiments. Recently, other decay processes of Xe atom, such as neutrino-less double electron capture ($0\nu\mathrm{ECEC}$) or neutrino-less quadruple beta decay~($0\nu4\beta^{-}$), are proposed to
theoretically explain lepton number violation~\cite{ecec, ecec2,quad}.

In these rare event search experiments, radioactive impurities in liquid
or gas Xe are severe background sources. In particular, the
noble gas radon-222~($\mathrm{^{222}Rn}$, or Rn)  is continuously produced from the decay of radium-226~($\mathrm{^{226}Ra}$) in the detector material.
Because of its long decay
time~(about $3.8$~days), the produced Rn enters into the sensitive
volume of the detector and results in generating mimic signals below
the multi-MeV region. Therefore, the removal of Rn in Xe is an essential
technique to improve the sensitivity of such experiments~\cite{xenon_nt, darwin}.

There are two strategies to reduce Rn in the detector: one is a careful
material screening before detector construction and the other is Rn removal during the observation phase with the detector. Before constructing the detector, radio impurity
of materials should be measured with a screening device such as a high-purity
germanium detector, an inductively coupled plasma mass
spectrometry~(ICP-MS), or any other special screening systems~\cite{exo_scan,
xenon100_scan, xenon_scan, panda_scan, next_scan, apc, aicham}.

Once the detector is constructed, the Rn emanation from any type of
material is basically fixed. Therefore, emanating Rn should be
removed through the purification process. Several studies for the Rn
removal in Xe gas have been conducted, for example, the development of
the distillation column and the single-column adsorption using an activated
charcoal~\cite{rn_xenon100, rn_xenon100_2, rn_xmass}.
Both Rn removal techniques have merits and demerits.
In the case of adsorption, a separation between Xe and Rn by using an adsorbent
is challenging because both elements are noble gases and have similar
molecule sizes.
Therefore, it is important to use appropriate adsorbent for the
selective adsorption of Rn in Xe. 
 
\subsection{Activated carbon fiber}

An activated charcoal is an effective adsorbent for various impurities by
physical process based on the van-del-Waals forces and polarizability
of the atoms~\cite{maurer}.
For example, an activated charcoal has an excellent ability to remove Rn from
argon~(Ar)~\cite{motoyasu}.
However, when the molecule sizes of the inert gas and Rn are similar,
such as Xe, both Rn and Xe are tended to be adsorbed by an adsorbent.
Therefore, the average pore size of the adsorbent should be selected carefully.
Another valuable property is an effective surface area of the adsorbent.
Larger surface areas allow increasing the adsorption efficiency.
Furthermore, since Rn emanation from the adsorbent itself ultimately
limits the Rn removal ability, it is required to reduce the amount of
internal radioactive impurities in the adsorbent.

Activated carbon fiber~(ACF) has been commercially available since the 1990s~\cite{acf0}.
ACF is a major adsorbent because it has a large surface area and good
adsorption property.
The general adsorption properties of ACF have been reported in various papers~\cite{acf1, acf2, acf3, acf4}.
However, its application to adsorb Rn in Xe gas has not been reported yet.

\begin{figure}[tb]
\centering\includegraphics[width=3in]{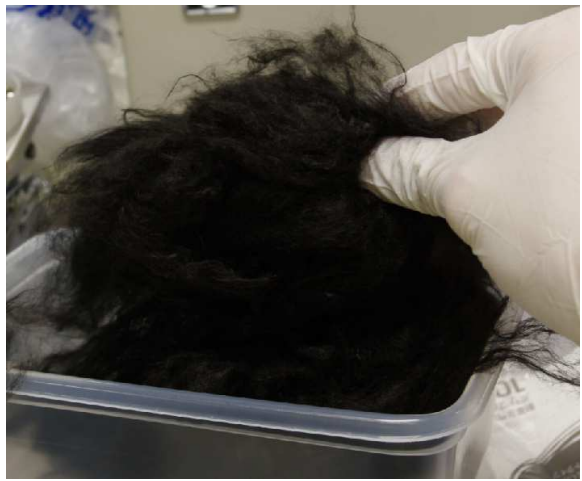}
\caption{A typical ACF provided from UNITIKA Ltd.}
\label{fig_fiber}
\end{figure}

For this study, several kinds of ACFs are specially provided from UNITIKA Ltd\footnote{\url{https://www.unitika.co.jp/e/index.html}}.
Figure~\ref{fig_fiber} shows a picture of a typical ACF provided from UNITIKA Ltd. 
The provided ACFs are named A-7, A-10, A-15, A-20, A-25 and S-25.
Among them, A-25 and S-25 are test products, and others are commercial products.
For the selective Rn adsorption, the pore parameters are essential. 
Such basic properties of these ACFs are summarized in Table~\ref{table_acf}.

\begin{table}[!h]
 \caption{Summary of basic properties of ACFs provided from UNITIKA Ltd.~(information provided from
 UNITIKA Ltd.). Meso pore volume ratio is the ratio of the pore volume of
 meso pores~($2$--$50$~nm) to the whole pore volume of an ACF.}
\label{table_acf}
\centering
\begin{tabular}{l||c|c|c|c|c|c}
\hline
                                & A-7  & A-10 & A-15 & A-20 & A-25 & S-25 \\ 
\hline \hline
Specific surface area [m$^2$/g] & $850$  & $1300$ & $1700$ & $2000$ & $2667$ & $1744$ \\ \hline
Average pore diameter [nm]      & $1.7$  & $1.7$  & $1.9$  & $2.2$  & $2.5$  & $5.3$  \\ \hline
Pore volume [cm$^3$/g]          & $0.35$ & $0.55$ & $0.80$ & $1.11$ & $1.65$ & $2.30$ \\ \hline
Meso pore volume ratio [\%]     & $4$    & $6$    & $10$   & $21$   & $37$   & $67$	  \\ \hline
\end{tabular}
\end{table}

The purpose of this report is to provide basic performances on Rn adsorption
of these ACFs in Xe in various conditions.
This report consists five sections, including this
introduction. In Section~\ref{sec_bg}, we describe the radioactive
impurities of those ACFs and compare their radioactivity with other activated charcoals reported in the earlier studies.
 In Section~\ref{sec_set_all}, we describe the
measurement system designed to evaluate the Rn adsorption efficiency of
the ACFs. Then, we present the performance of this system using purified
air~(Air), Ar gas, and Xe gas.
In Section~\ref{section_result}, 
we present the characteristic dependence of Rn adsorption in Xe, and
show the comparisons of the adsorption efficiency among those ACFs. 
In the last section, we summarize this study.

\section{Radioactive impurity measurements} \label{sec_bg}

As explained in Section~\ref{sec_intro}, the radon emanation from an
adsorbent limits the Rn adsorption ability. 
Therefore low-intrinsic radioactivity of $\mathrm{^{226}Ra}$ (or uranium
series) is required for an adsorbent.
We measured the intrinsic radioactive contamination of the provided
ACFs with high-purity germanium detectors~(HPGe) in Kamioka Observatory, the Institute for Cosmic Ray Research~(ICRR), 
the University of Tokyo and in Kavli Institute for the Physics and Mathematics of the Universe~(IPMU), the University of Tokyo.
For this evaluation, we measured four ACFs, which are A-10,
A-15, A-20, and A-25.
We put those ACFs into an EVOH~(Ethylene-vinylalcohol copolymer) plastic
bag in order to keep emanated Rn from the samples, then the measurements
are carried out for about $10$~days after reaching radiation equilibrium.
This HPGe assay method is reported in Ref.~\cite{HPGe}. 

The results of the measurements are summarized in
Table~\ref{table_purity} as well as the other measurements of activated
charcoals by the XMASS collaboration~\cite{rn_xmass} and the XENON collaboration~\cite{loss}.
\begin{table}[!h]
\caption{Summary of intrinsic radioactivities of A-10, A-15, A-20 and
 A-25. Measurements of activated charcoals are also shown.
 The measurement device is a high-purity germanium detector~(HPGe) or
 proportional counter~(PC).
 For the HPGe measurements, lead-214~($\mathrm{^{214}Pb}$) and bismuth-214~($\mathrm{^{214}Bi}$)
 are used for the estimation of uranium series, and
 actinium-228~($\mathrm{^{228}Ac}$) is used for the estimation of thorium series.
 The upper limit values are $90\%$ confidence level.}
\label{table_purity}
\centering
\begin{tabular}{c||c|c|c|c|c|c}
\hline
 & Method & Weight & Duration & Uranium & Thorium & $\mathrm{^{40}K}$ \\
 &        &        &          & series  & series  &  \\ 
 & & [g] & [day] & [$\mathrm{mBq/kg}$] & [$\mathrm{mBq/kg}$] & [$\mathrm{mBq/kg}$] \\
\hline \hline
A-10 & HPGe & $6.65$ & $12.4$ & $<352$ &$<305$ & $<4.31\times10^{3}$\\ \hline
A-15 & HPGe & $160.0$ & $10.1$ & $<11.9$ &$<12.2$ & $<142$\\ \hline
A-20 & HPGe & $267.4$ & $11.6$ & $<5.5$ &$<10.4$ & $<49$\\ \hline
A-25 & HPGe & $8.4$ & $14.9$ & $<269$ &$<261$ & $<4.31\times10^{3}$\\ \hline
Shirasagi G$_{\mathrm{2X}}$~$4/6$~\cite{rn_xmass} & HPGe& $95.0$ & $7.0$ & $67\pm15$ & -- & -- \\ \hline
Shirasagi G$_{\mathrm{2X}}$~$4/6$~\cite{loss} & PC & -- & -- & $62\pm4$ & -- & -- \\ \hline
Bl$\mathrm{\ddot{u}}$echer 100050~\cite{loss} & PC & -- & -- & $2.6\pm0.3$ & -- & -- \\ \hline
\end{tabular}
\end{table}
For these ACFs, we could not observe any significant gamma-ray line over the background spectrum in the
measurements in Table~\ref{table_purity}\footnote{
However, we observed $63 \pm 9 ~\mathrm{mBq/kg}$ of $\mathrm{^{228}Th}$
(half life is $1.9$~years) only in the A-20 sample.
We think it would be a contamination in the preparation process of this A-20
sample, and this contamination will not affect $\mathrm{^{222}Rn}$ assays.}.
Therefore, we have estimated the upper limits.
The obtained upper limit of A-20 is
$<5.5$~$\mathrm{mBq/kg}$ for the uranium series.
Comparing the radioactivity of the uranium series with other
reports~\cite{rn_xmass,loss}, A-20 has a lower radioactivity than Shirasagi G$_{\mathrm{2X}}$~$4/6$.
However, we could not judge whether A-20 has a lower~(or higher)
radioactivity than Bl$\mathrm{\ddot{u}}$echer 100050
\footnote{
There is a difference of the estimation.
PC measures emanated radon from the sample only, but
HPGe measures radioisotopes inside the sample also.
}.

\section{Experimental setup} \label{sec_set_all}

\subsection{Test measurements with purified air and argon} \label{sec_test}

As a first step, we evaluated the Rn adsorption performance of ACF in
purified Air and in purified Ar.
We used G1 grade purified gases (impurities are less than $0.3$ ppm)
provided from TAIYO NIPPON SANSO CORPORATION~\cite{G1gas}.
Since the molecular sizes of nitrogen~(N$_2$), oxygen~(O$_2$) and Ar
are relatively small compared with that of Rn, the adsorption of the Rn
in purified Air or Ar is expected to be efficient.
To date, adsorption techniques of Rn in Air or N$_2$ with activated charcoals have been established among underground
experiments~\cite{ynakano, sk, sno, nemo, borex}. 

In order to perform the test measurements with Air or Ar, we have constructed a
test bench at Kobe University, Japan. The test bench is based on the
measurement system developed in Ref.~\cite{motoyasu}.
The test bench for this study consists of a high-sensitivity 80-L Rn
detector~\cite{hosokawa,ynakano},
a gas mass flow controller~(HORIBA STEC, SEC-Z512MGX, hereafter MFC), a gas
circulation pump~(ENOMOTO Micro Pump Mfg. Co. Ltd., MX-808ST-S), a dew
point gauge~(VAISALA, DMT152), a pressure gauge~(Swagelok,
PGU-50-MC01-L-4FSF), filters~(NIPPON SEISEN Co. Ltd., NASclean GF-T001
and GF-D03N), a refrigerator with a cold trap, and a Rn
source (PYLON, RNC).  The radioactivity of this source is $78.3$~Bq~($\mathrm{^{226}Ra}$).
This refrigerator is equipped with a box-shaped cold trap~\cite{motoyasu}.
A schematic view of this test bench corresponds to the system shown
in Figure~\ref{fig_system} in Section~\ref{sec_setup} without
the main trap in that figure.  

The principal techniques of the 80-L Rn detector are the electrostatic
collection of the positively charged daughter nuclei of
$\mathrm{^{222}Rn}$~\cite{positive1, positive2} and the deposit energy
measurement of their $\alpha$ decays on a PIN 
photodiode~\cite{rn1,rn2,rn3,rn4, hosokawa, ynakano}. The 80-L Rn
detector can measure a few $\mathrm{mBq/m^{3}}$ level of Rn
concentration in the circulation gas.
The details of the 80-L Rn detector can be found in Ref.~\cite{hosokawa, ynakano}. 

Here are typical operations before the test measurements with this test bench.
For the measurements in this section, we put $4.75$~g of ACF A-15 into the box-shaped trap. 
After preparation of the ACF sample in a box-shaped trap,
we baked the trap at $+85 \mathrm{^{\circ}C}$ under vacuum with a
turbo molecular pump.
After the vacuum pressure became  less than $1.0\times10^{-1}$~Pa, we stopped the
baking, then started cooling down of the box-shaped trap to $-105 \mathrm{^{\circ}C}$.
In parallel, the test bench was filled with purified Air or Ar. 
When we changed the gas in the system, the entire system was evacuated
before filling, except for the Rn source.
The Rn source was purged with the new gas in advance.
The typical pressure of the gas in the system is at atmospheric pressure
(around $0.10$~MPa in absolute pressure, or $\pm 0.000$~MPa in gauge pressure).
It is noted in this report that we use gauge pressure values to express the inner gas pressure of the test bench
if it is not specified. The gas was circulated with the circulation pump after filling.

There are three steps after the preparation of the cold trap and
the circulation gas.
The first phase is the injection of Rn into the circulation gas from the Rn source.
The second phase is the adsorption of Rn with the cooled box-shaped trap.
The third phase is the release of Rn from the trap by heating the trap
to the normal temperature (between $+15 \mathrm{^{\circ}C}$ and $+25 \mathrm{^{\circ}C}$).
During these phases, the flow rate of the circulation gas was kept at the
same flow rate with the MFC in the system.
The 80-L Rn detector was also continuously monitoring the Rn concentration
in the circulation gas, during these phases.
The observed number of polonium-214~($\mathrm{^{214}Po}$) events from the 80-L Rn detector was
summarized every 10 minutes, then the observed count rate is converted
into the Rn concentration with the calibration factor in Ref.~\cite{hosokawa,ynakano}.
Therefore, we obtained Rn concentration in the circulation gas every 10 minutes.

\begin{figure}[tb]
\begin{minipage}{0.5\hsize}
\centering\includegraphics[width=3in]{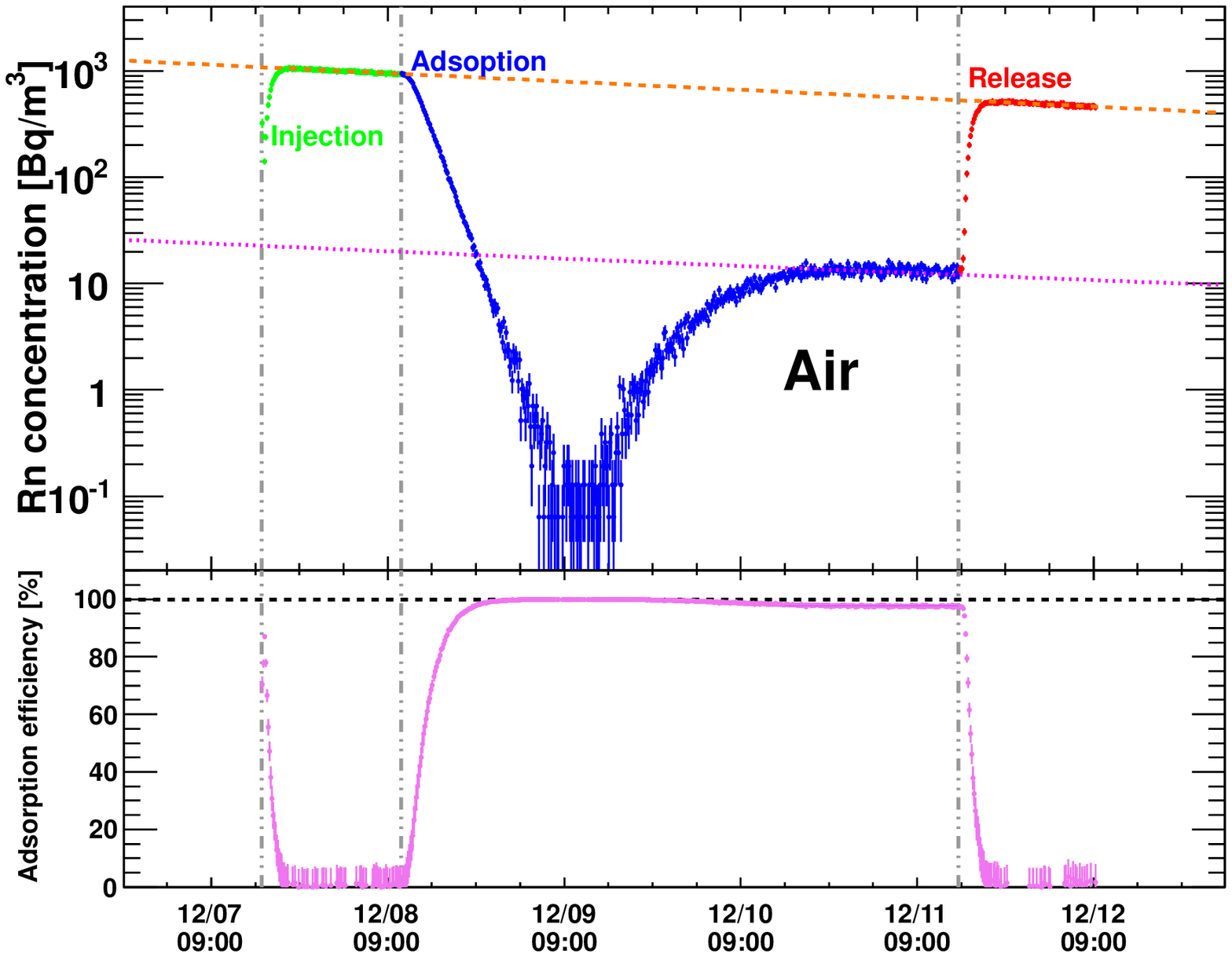}
\end{minipage}
\begin{minipage}{0.5\hsize}
\centering\includegraphics[width=3in]{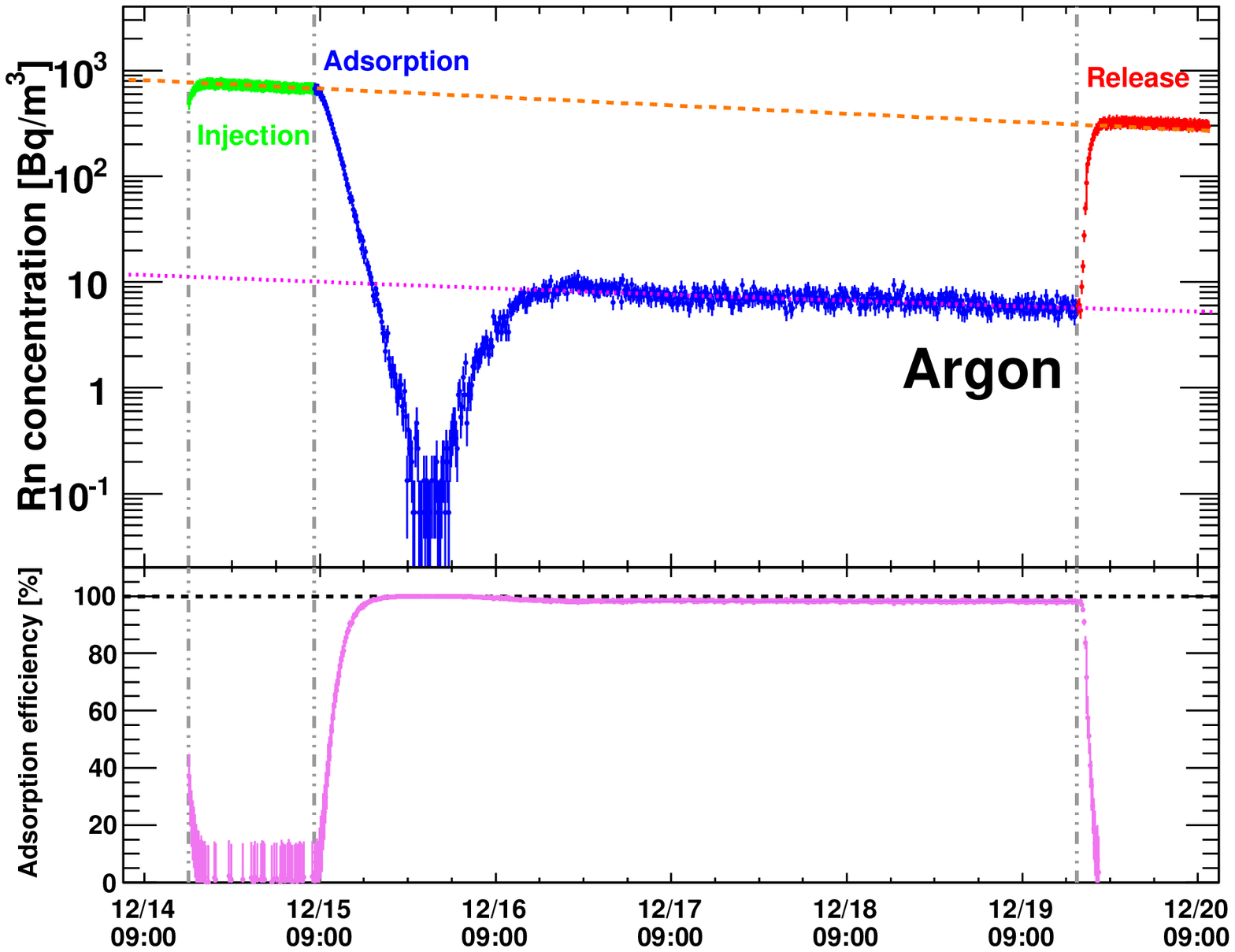}
\end{minipage}
\caption{The left side (right side) plot shows measurements with
 purified Air~(Ar).
 The upper~(lower) panel shows the Rn concentration~(Rn adsorption efficiency) as a function of time.
In the top panel, the colored data points show the different phases of the measurements, where green,
 blue, and red show the Rn concentration in the injection phase, in the adsorption phase, and the release phase, respectively. 
 The orange~(pink) line in the upper panel shows the
 Rn decay curves for expected~(remaining) Rn concentrations.}
\label{fig_air}
\end{figure}

Figure~\ref{fig_air} shows the results of these test measurements.
For the test measurement with Air, we injected Rn at 15:51 on December 7th,
2016. The gas circulation rate was $0.90$~SLM (liter per minute at standard temperature and pressure) and then we started the
adsorption phase from 10:52 on December 8th. The Rn concentration
dropped soon after starting the adsorption phase and it took about
$24$~hours to reach the lowest value. At the moment, the Rn
adsorption efficiency reached a maximum of nearly $100\%$ with the reduction factor of $10^{4}$. 

In general, further adsorption is not expected when all pore sites are
filled with gas molecules. Under such situation,
adsorbed Rn is eventually released from ACF by collision with other molecules in circulation gas. Consequently,
some of Rn returns into the circulation gas again because of such
saturation.
This situation is called break-through.
The Rn concentration in the system ultimately reaches the equilibrium state
between the adsorption and the release. 

In the  measurements with Air or Ar, the Rn concentration gradually increased to about
${\sim}10~\mathrm{Bq/m^{3}}$ level after reaching the lowest value, and this can be explained as the reason above.
During this phase, the Rn
adsorption efficiency finally becomes $97.9\pm0.1\%$. 
After starting the release phase at 14:39 on December 11th, the Rn
concentration immediately increased and then reached to the originally
expected level. 

For the test measurement with Ar, we have evaluated the Rn adsorption efficiency
with the same procedure with the gas circulation rate at $1.3$~SLM.
The result is shown in Figure~\ref{fig_air}~(right).
Its Rn adsorption efficiency is determined to be $98.3\pm0.1\%$ as
summarized in Table~\ref{table_air_argon}.
\begin{table}[tb]
\caption{Summary of the Rn adsorption efficiency in the purified Air and 
 Ar with A-15. The definition of the Rn adsorption efficiency is explained in
 Section~\ref{sec_ana}.
 The possible systematic uncertainties are
 Pressure drop in the adsorption phase $-2.0\%$,
 Accuracy of pressure $-1.0\%$, and
 Reproducibility $\pm 2.0\%$.
 }
\label{table_air_argon}
\centering
\begin{tabular}{c||c|c}
\hline
Gas & Flow rate~[SLM] & Rn adsorption efficiency~[$\%$]\\ 
\hline \hline
Air  & 0.90 & $97.9\pm 0.1$ (stat.) \\ \hline
Argon & 1.3 & $98.3\pm 0.1$ (stat.) \\ \hline
\end{tabular}
\end{table}
Though the adsorption efficiency of Air and Ar measurements are similar,
the break-through time of Ar looks shorter than that of Air.
This would be due to the difference of the gas flow rates of these tests.
From these test measurements, we demonstrated that this system can precisely
measure the Rn adsorption efficiency of ACF in both Air and Ar. In addition to this,
we found that ACF A-15 showed a good Rn adsorption efficiency.

\subsection{Measurement system for Xe gas} \label{sec_setup}

\begin{figure}[tb]
\centering\includegraphics[width=5in]{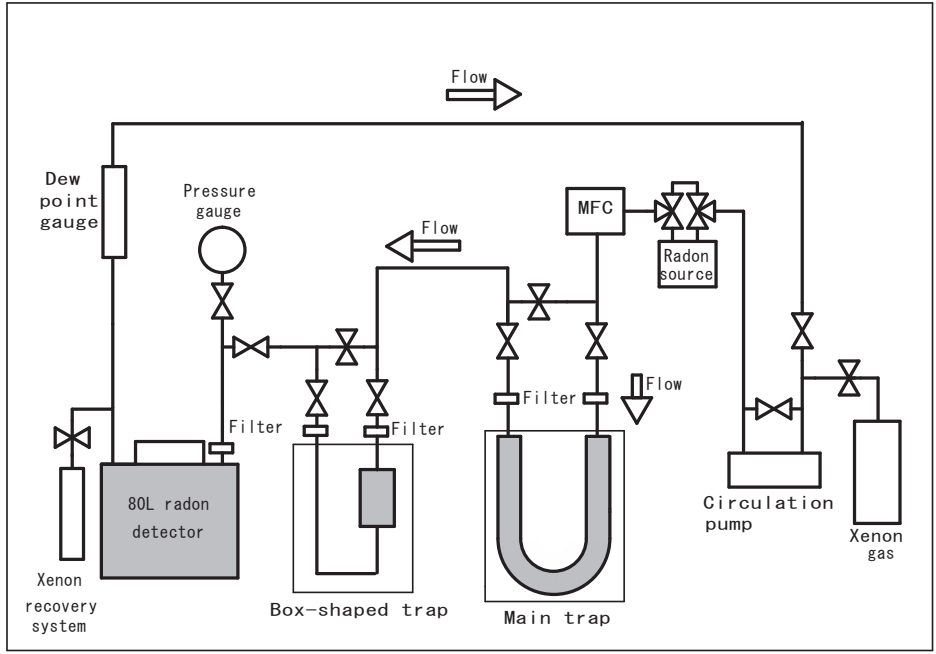}
\caption{Experimental setup of the full test bench. The Rn detector
 was upgraded to an 80-L detector from the system in
 Ref.~\cite{motoyasu}.
 Then, a new refrigerator and a new larger cold trap~(main trap) were
 newly added for the measurements with Xe gas. The arrows show the direction of the circulation gas flow.}
\label{fig_system}
\end{figure}

For the measurements using Xe gas, we
added a new refrigerator with a larger cold trap to increase the amount
of the ACF sample.
A schematic diagram of the full system is shown in Figure~\ref{fig_system}.
The gas in the system was circulated by the circulation pump.
The Rn concentration in the circulation gas was measured with the 80-L Rn detector.
The pressure, temperature, and dew point were monitored continuously during the measurement.

\begin{figure}[tb]
\centering\includegraphics[width=3in]{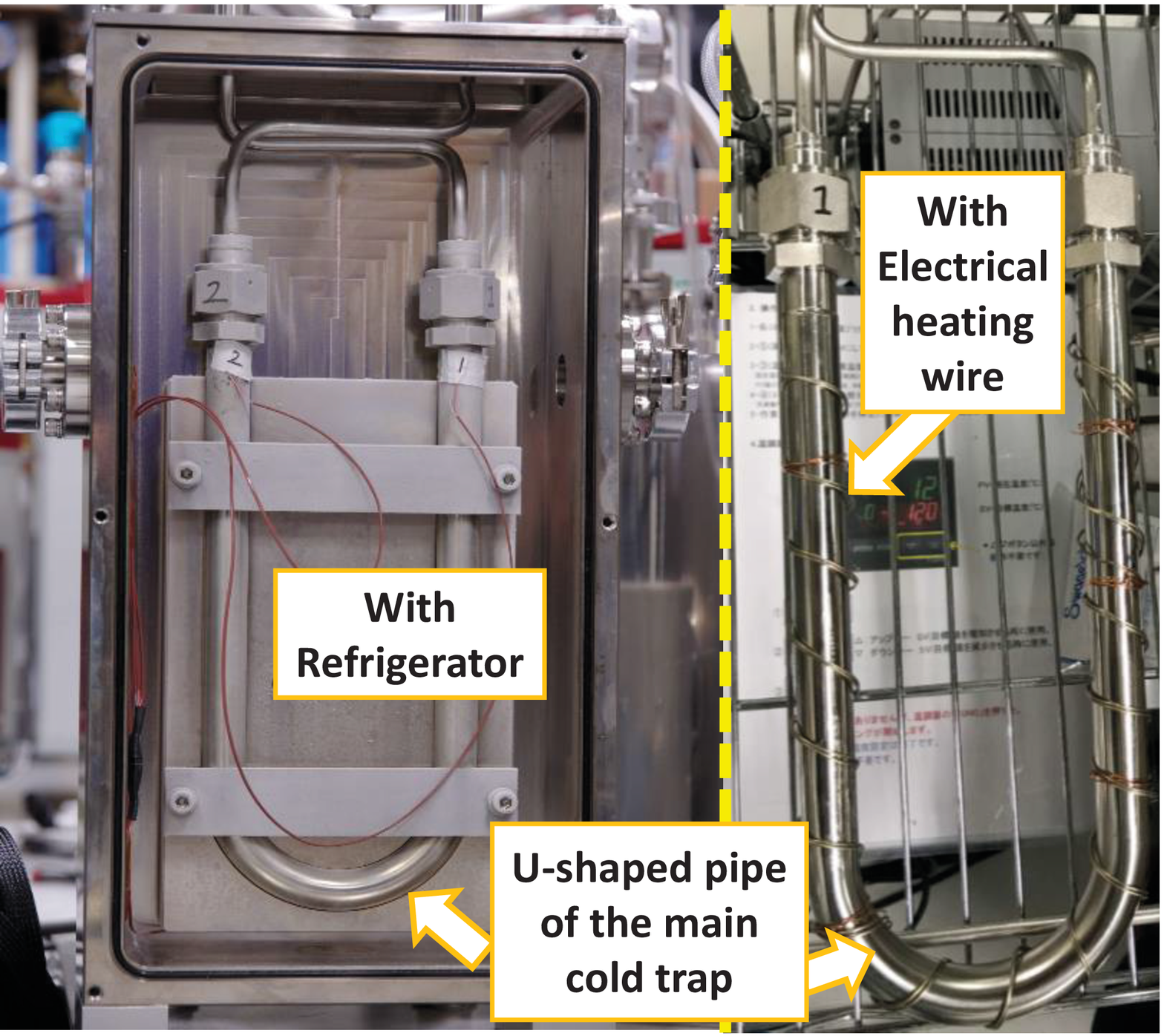}
 \caption{The photographs of the main trap. The left side shows
 the trap in the cooling system for the adsorption phase, and the right
 side shows the trap with 
 an electrical heating wire for baking in the release phase.}
\label{fig_trap}
\end{figure}

In this system, two cold traps~(main trap and box-shaped
trap) were used.
The main trap is a U-shaped electro-polished stainless steel pipe, and
can typically hold  $10$--$20$~g of an ACF sample.
The main trap is used to hold the ACF sample to estimate Rn adsorption
efficiency, and its volume is 122 cm$^3$.
The temperature of the ACF sample was controlled by the refrigerator
for the main trap.
Photographs of the main trap are shown in Figure~\ref{fig_trap}.
The box-shaped trap is used to control humidity in the circulation gas for the measurements with Xe.
In the actual measurement with Xe gas, the box-shaped trap was always
kept at $-70 \mathrm{^{\circ}C}$ to keep low humidity conditions.
We tested this box-shaped trap in Xe gas, then confirmed the effect
from the box-shaped trap was negligible in Xe gas.  

In order to prevent possible contamination of carbon fibers from the ACF
sample into the measurement system, thin-layer metal membrane filters
($0.0025~\mu$m and $0.3~\mu$m) are installed at the inlet and outlet of the
cold traps. 

\subsection{Measurement method} \label{sec_method}

Here is a typical measurement method with the main trap.
Before the measurements, we put an ACF sample into the main trap. 
Then we evacuated the main trap with baking at ${+}180 \mathrm{^{\circ}C}$.
After the vacuum pressure became less than $1.0\times10^{-1}$~Pa in absolute pressure,
 we stopped the baking, and then put the main trap into the cooling system.
Finally, the measurement system was filled with Xe gas.
The typical circulation gas pressure was at atmospheric pressure, but it
was reduced down to $-0.071$~MPa in the case of the pressure dependence measurement described in Section~\ref{sec_press}.
The Xe gas was circulated with the pump after filling at $0.14$--$1.4$~SLM.

The three steps (injection, adsorption, and release phases) are same as the test measurement described in Section~\ref{sec_test},
but the temperature condition was changed for the main trap. 
All the temperature setting of the main refrigerator during the adsorption
phase are $-95 \mathrm{^{\circ}C}$.
The typical temperature setting during the release phase is normal
temperature. In some measurements, the temperature at the beginning of
the release phase was set at $+180 \mathrm{^{\circ}C}$ to increase
release speed. 
In each step, the measurement continues until the count rate of the Rn
detector becomes stable. 
The adsorption phase was started by switching the inlet and outlet 
valves of the main trap from by-pass mode to through-trap mode.
Using the 80-L Rn detector, the Rn concentration in the circulation gas
was continuously monitored.

\subsection{Data analysis} \label{sec_ana}

In order to measure the Rn adsorption efficiency in the circulation gas,
we have performed the following analysis procedure.
At first, the Rn concentration in the circulation gas is determined by
the fitting with the Rn decay curve, defined as
$C_{0}e^{-\lambda t}+C_{1}$~[$\mathrm{Bq/m^{3}}$], where $\lambda$ is
the decay constant of $^{222}$Rn, $t$ is elapsed time, $C_{0}$ and $C_{1}$ are
parameters to be fitted.
Applying this equation to the data before the adsorption phase,
the expected Rn concentration without Rn adsorption ($C_{\mathrm{expected}}$) at
$t = 0$ is obtained as $C_{0} + C_{1}~\mathrm{[Bq/m^{3}]}$.
When this equation is applied to the data during the adsorption phase,
the remaining Rn concentration after Rn adsorption by the cold
trap ($C_{\mathrm{remaining}}$) at $t = 0$ is obtained, as well.
The expected fitted line is compared with the data after the release phase, in
order to check the consistency of the $C_{\mathrm{expected}}$ value.

Then, we compare the remaining Rn concentration with the expected Rn
concentration as follows; 
\begin{equation}
 R = \frac{ C_{\mathrm{expected}} - C_{\mathrm{remaining}} } { C_{\mathrm{expected}} } \times 100.0~[\%].
\end{equation}
In this report, we use the $R$~[$\%$] as the Rn adsorption efficiency.

\subsection{A typical measurement in Xe gas and systematic uncertainty} \label{sec_syst}

\begin{figure}[tb]
\begin{minipage}{0.5\hsize}
\centering\includegraphics[width=3in]{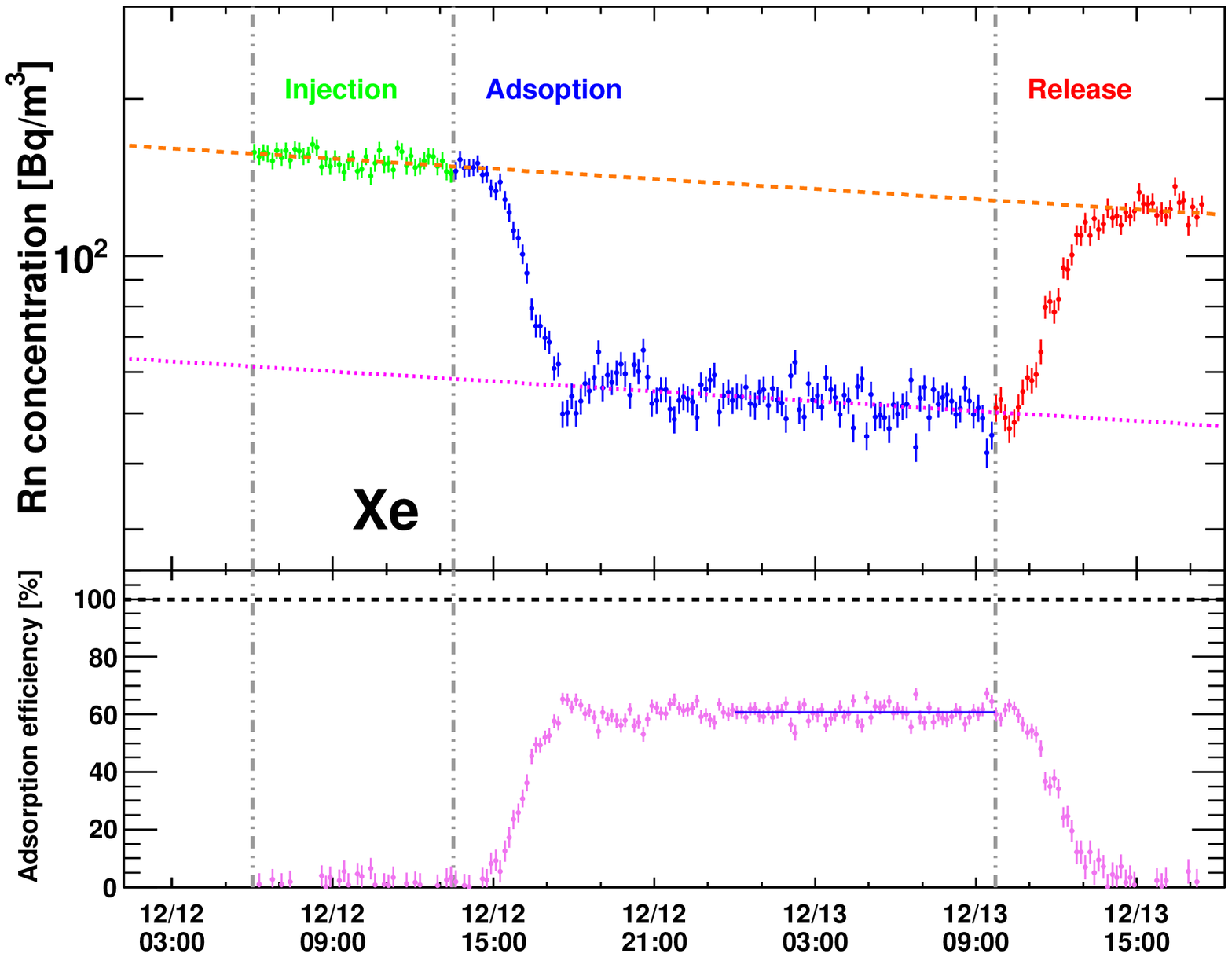}
\end{minipage}
\begin{minipage}{0.5\hsize}
\centering\includegraphics[width=3in]{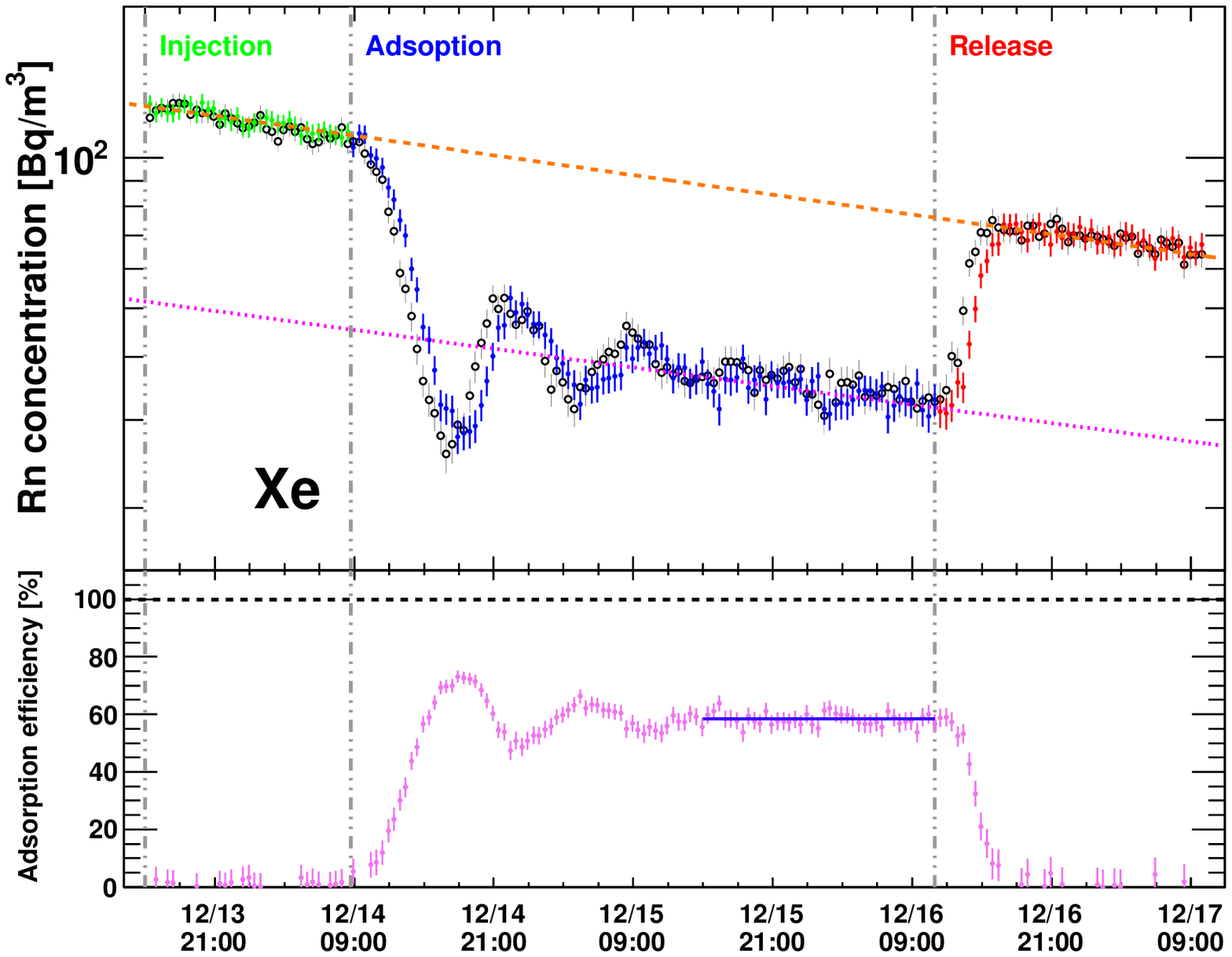}
\end{minipage}
 \caption{An example of the measurements with Xe gas with ACF A-25.
 The Xe gas pressure before the adsorption phase,
 Xe gas pressure before the release phase, and flow rate in left~(right)
 are $-0.071$~MPa~($-0.071$~MPa), $-0.076$~MPa~($-0.075$~MPa) and
 $0.41$~SLM~($0.14$~SLM), respectively. 
 The plot and color definitions are the same as Figure~\ref{fig_air}.
 The black open circles in the right-upper plot is Rn concentration
 obtained from $^{218}$Po counts.   
 }
 \label{fig_xe1}
\end{figure}

Figure~\ref{fig_xe1} shows an example of the measurements with Xe gas.
In these measurements, ACF A-25 was used in Xe circulation gas at
$-0.071$~MPa, and we observed a clear drop of the Rn concentration during the
adsorption phase. The Rn concentration drops in $0.41$~SLM and $0.14$~SLM
flow rate measurements are very similar at around $60\%$. 
An apparent oscillation of the Rn concentration in the adsorption phase is
observed in the $0.14$~SLM flow rate measurement.
This behavior is similar to Ref.~\cite{rn_xmass}.
  
We also observed a pressure difference between the adsorption
phase and other phases.
The pressure drop during the adsorption phase was due to  the adsorption of  both Xe and Rn  by the ACF. 
This pressure difference suggests an actual Rn adsorption efficiency might be lower than observed
$R$ defined in Section~\ref{sec_ana}.
Therefore we have evaluated the amount of the pressure difference
as a systematic uncertainty on the Rn adsorption efficiency.

In the case of the $0.41$~SLM measurement, the pressure difference was $0.0050$~MPa and the
absolute pressure before the adsorption phase was $0.0303$~MPa. Based on this difference, 
we assigned the systematic error of $-16.5\%$ on the absorption efficiency.
Typical gauge pressure values of the circulation gas were $\pm0.000$~MPa
in the adsorption phase and $+0.005$~MPa in other phases, except for the pressure dependence measurement.
This pressure difference  corresponds to $-5.0\%$ systematic uncertainty, typically.

The accuracy of the pressure of the circulation gas has around
$\pm 0.001$~MPa uncertainty, since the pressure gauge is an analog type.
Because of this accuracy, we assigned a few~$\%$ level of uncertainty on the Rn adsorption efficiency.
This systematic uncertainty ranges from $-3.0\%$ to $-1.0\%$ depending on the pressure
inside the system~($-3.0\%$ for the lowest case of $-0.07$~MPa and $-1.0\%$ for atmospheric pressure).
This uncertainty will reduce or enhance the uncertainty
on the Rn adsorption efficiency from the pressure drop.
We estimated maximum enhancement as a systematic error source.

In the very similar condition measurements, the observed Rn adsorption
efficiency was slightly different within about a few~$\%$.
Therefore, we assigned systematic uncertainty from reproducibility as $\pm2.0\%$.

The assigned systematic uncertainties in this study are summarized in 
Table~\ref{table_systematic}.
\begin{table}[!h]
\caption{Summary of the systematic uncertainties on the Rn adsorption efficiency.}
\label{table_systematic}
\centering
\begin{tabular}{l|c}
\hline
Source                 & Uncertainty\\ 
\hline \hline
Pressure drop in the adsorption phase & $-5.0\%$~(typical) \\ \hline
Accuracy of pressure          & Between $-3.0\%$~(at $-0.07$~\rm{MPa}) and $-1.0\%$~(at 0.00~\rm{MPa}) \\ \hline
Reproducibility            & $\pm 2.0\%$ \\ \hline
\end{tabular}
\end{table}

The observed Rn adsorption efficiency values in Xe of the $0.41$~SLM
and $0.14$~SLM flow rate measurements were
$60.8 \pm0.3$(stat.)$^{+2.0}_{-16.9}$(syst.)\%, and 
$58.5 \pm0.3$(stat.)$^{+2.0}_{-13.6}$(syst.)\%, respectively. 
The uncertainty from the pressure drop in the adsorption phase is estimated
with actual pressure drop in each measurement.

In preceding studies~\cite{rn_xmass, loss}, the velocity ratio
(or retention time ratio) of Rn to Xe in the adsorbent is estimated
as a Rn trap performance of the test system.
We also estimate the velocity ratio from the measurement in
Fig.~\ref{fig_xe1} right.
In this measurement, we have observed retention time of Rn
($T_{\rm{Rn}}$ in Ref.~\cite{rn_xmass}) in the
main trap as $10 \pm 2$ hours, using the $^{218}$Po data,
then the velocity ratio becomes $v_{\rm{Rn}} / v_{\rm{Xe}} = (1.4 \pm
0.2) \times 10^{-3}$ (at $-95 \mathrm{^{\circ}C}$).
The corresponding ratios in the preceding studies are
$v_{\rm{Rn}} / v_{\rm{Xe}} = (0.96 \pm 0.10) \times10^{-3}$ (at $-85 \mathrm{^{\circ}C}$)~\cite{rn_xmass}
and
$v_{\rm{Rn}} / v_{\rm{Xe}} = (0.65 \pm 0.11) \times 10^{-3}$ (at $-70 \mathrm{^{\circ}C}$)~\cite{loss}.
This means velocity of Rn in the adsorbent column in our test bench
(= main trap) is faster than others. 
The main reason of this difference would be due to the difference
of the density of the adsorbent in the column.
From the information in Ref.~\cite{rn_xmass}, the adsorbent density in the
column is estimated as about 358 kg/m$^3$.
On the other hand, that from our system is 98 kg/m$^3$, since
we could not pack fibers into the U-shaped column very efficiently.

Then, using this velocity ratio, we also estimate the two component 
Henry's constant of the adsorbent $\tilde{H}$ which is introduced in
Ref.~\cite{loss}.
In the report, they obtained $\tilde{H} = (1.94 \pm 0.24) \times
10^{-3}$ mol/(Pa kg) (at $-70 \mathrm{^{\circ}C}$, $-0.080$ MPa Xe).
From our measurement,
we obtained $\tilde{H} = (5.0 \pm 1.0) \times 10^{-3}$ mol/(Pa kg)
(at $-95 \mathrm{^{\circ}C}$, $-0.071$ MPa Xe).
Though the conditions of the measurement are different, our adsorbent
shows comparable adsorption performance with the preceding work.

 \section{Result} \label{section_result}

In this section, we describe the measurement results of the Rn
adsorption efficiency in Xe gas. As a first step, we evaluated a possible
flow rate dependence of the Rn adsorption efficiency using ACF A-25.
Then, its Rn concentration dependence and circulation gas pressure
dependence was evaluated with A-25. After presenting the results from these dependence measurement,
we discuss the Rn adsorption efficiency among the six ACFs provided by the company.

\subsection{Flow rate dependence}

At first, we evaluated the flow rate dependence of the Rn adsorption
efficiency using
$10.50$~g of A-25.
The experimental procedure is the
same as the typical measurement explained in Section~\ref{sec_method},
but we changed the circulation gas flow rate using MFC. 
We also kept the Rn source connected during this measurement in order to maintain the same Rn concentration level. 
Therefore, the difference should be only the flow rate of the circulation gas. 
Typical Rn concentration and circulation gas pressure are
$10^{3}~\mathrm{Bq/m^{3}}$ and atmospheric pressure, respectively.

The obtained results are summarized in Table~\ref{table_flow}. 
\begin{table}[!h]
\caption{Summary of the flow rate dependence of the adsorption efficiency with A-25.}
\label{table_flow}
\centering
\begin{tabular}{c|c}
\hline
Flow rate [SLM] & Rn adsorption efficiency~[$\%$]\\ 
\hline \hline
$0.14$ & $27.8 \pm 0.2$(stat.)$^{+2.0}_{-5.5}$(syst.) \\ \hline
$1.4$ & $27.4 \pm 0.4$(stat.)$^{+2.0}_{-5.5}$(syst.) \\ \hline
\end{tabular}
\end{table}
Here, the systematic errors are obtained from Table~\ref{table_systematic}. However, for the
relative comparison in this estimation, the relevant systematic
uncertainty would be only reproducibility ($\pm2.0\%$), since the pressure
drops are very similar in these measurements.

Among these different flow rate measurements, no significant difference was observed. 
Therefore, we found no significant flow rate dependence of measured Rn
adsorption efficiency in the test bench in between $0.14$ and $1.4$~SLM flow rate. 

\subsection{Rn concentration dependence}

As explained in Section~\ref{sec_test}, further adsorption is not
expected when all pore sites are filled with gas molecules.
Therefore, the Rn adsorption efficiency may depend on the Rn concentration
of the circulation gas.
In order to evaluate this dependence, we measured the
Rn adsorption efficiency under the different Rn concentrations in the
circulation gas. In these measurements, $10.50$~g of A-25 was used.
Typical flow rate and circulation gas pressure are 1.4~SLM and
atmospheric pressure, respectively. 

\begin{figure}[tb]
\begin{minipage}{0.5\hsize}
\centering\includegraphics[width=3.2in]{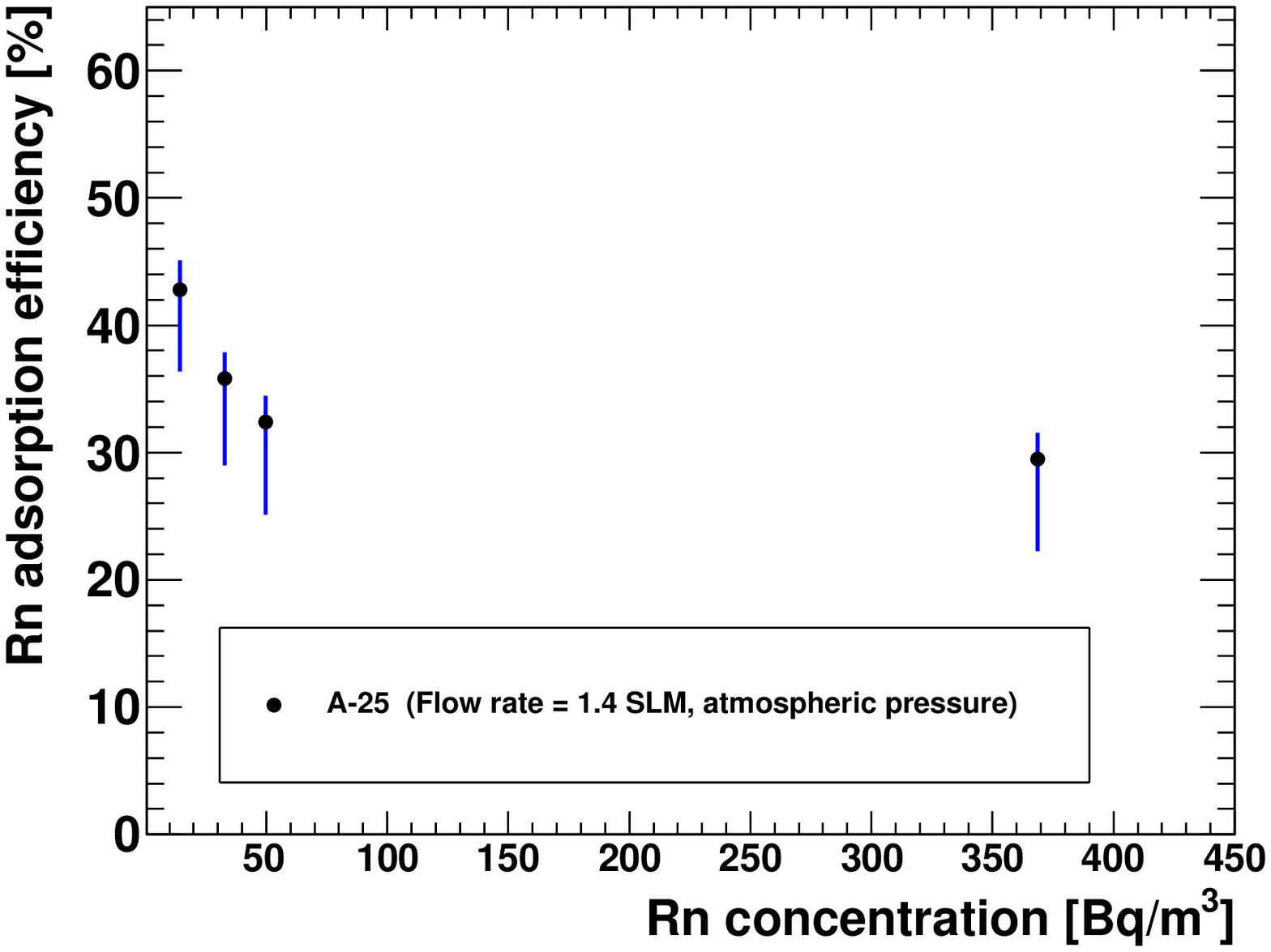}
\end{minipage}
\begin{minipage}{0.5\hsize}
\centering\includegraphics[width=3.2in]{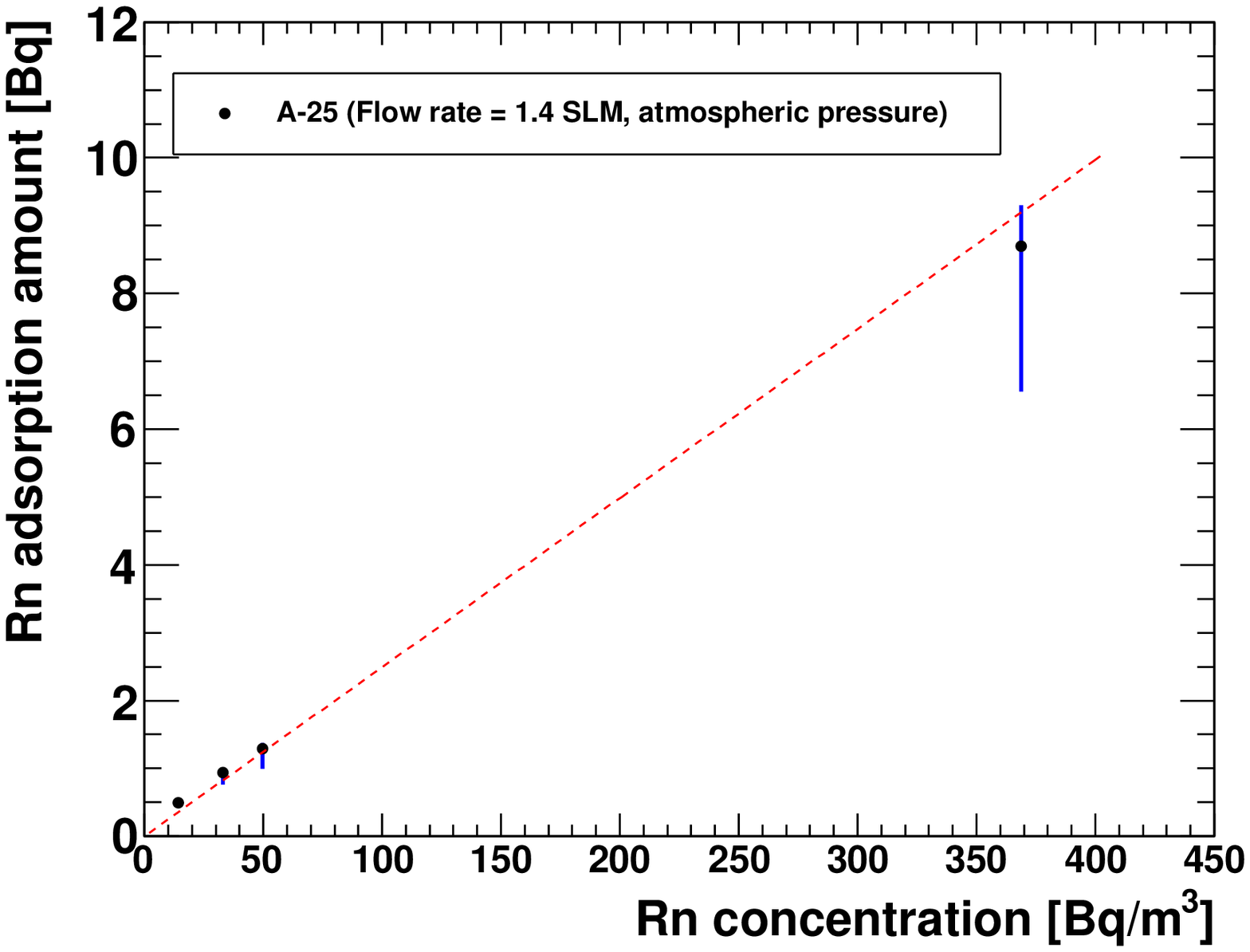}
\end{minipage}
 \caption{
 The left side (right side) plot shows 
 Rn adsorption efficiency (Rn adsorption amount) as a function of the Rn
 concentration just before the adsorption phase.
 The black points are observed $R$, and the blue bars are statistical and systematical errors, added in quadrature.
 The red dashed line in the right side plot is a fitting with a straight line
 ($(Rn \ adsorption \ amount) = (0.025 \pm 0.001) \times (Rn \ concentration)$).
 We used $10.50$~g of A-25 in these measurements. The circulation gas pressure of all the data points is atmospheric pressure.
 The flow rate of the circulation gas is fixed at $1.4$~SLM.
 The temperature setting of the main refrigerator during the adsorption
 phase is fixed at $-95 \mathrm{^{\circ}C}$.
 }
\label{fig_con_com}
\end{figure}

Figure~\ref{fig_con_com} shows the measurement results.
From Fig.~\ref{fig_con_com} right, the absorbed Rn amount looks
generally proportional to the Rn concentration.
Since the Rn concentration corresponds to the partial pressure of Rn
gas, this measurement may indicate Henry's law in Rn adsorption~\cite{loss}. 

However, from Fig.~\ref{fig_con_com} left, the Rn adsorption efficiency
looks increasing in the region below $50~\mathrm{Bq/m^{3}}$.
This dependence suggests that a high Rn absorption efficiency is
expected in the real experiments which search for rare physics events
because the Rn concentration level in such experiments is much less than
this Rn concentration level.
In the region of $50$--$300~\mathrm{Bq/m^{3}}$, the Rn concentration dependence of
the Rn adsorption efficiency is about $\pm3\%$, and this is similar as or slightly smaller than
the systematic uncertainty of the Rn adsorption efficiency estimated in
Section~\ref{sec_syst}.

\subsection{Pressure dependence} \label{sec_press}
\begin{figure}[tb]
\centering\includegraphics[width=4in]{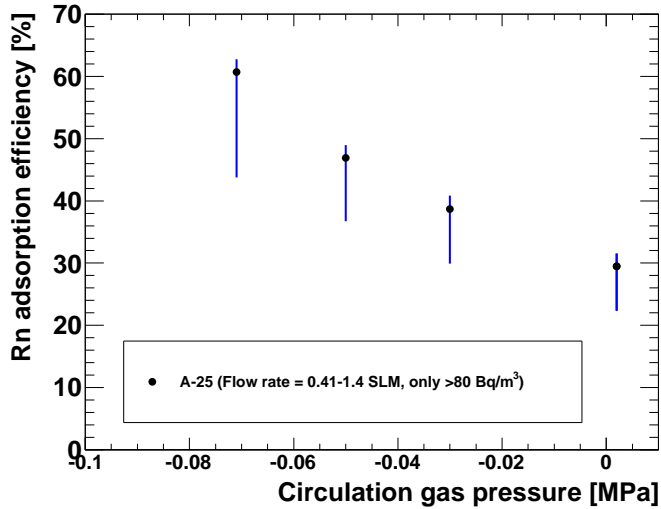}
\caption{Rn adsorption efficiency as a function of circulation gas
 pressure. The black points are observed $R$, and the blue bars are
 statistical and systematical errors, added in quadrature.
 $10.50$~g of A-25 was used. The Rn concentration of all the
 data points are above 80 Bq/m$^3$.
 The flow rate is $1.4$~SLM for the atmospheric pressure measurement
 and $0.41$~SLM for others.
 The temperature setting of the main refrigerator during the adsorption
 phase is fixed at $-95 \mathrm{^{\circ}C}$.
 } 
\label{fig_press_dep}
\end{figure}

In general, an adsorption on porous depends on both the temperature and the amount of the adsorbed gas molecules. Therefore, the
 adsorption may depend on the circulation gas pressure inside the system.
In order to evaluate the dependence on the circulation gas pressure, we measured the Rn adsorption efficiency of A-25 under the different circulation
gas pressure. For this measurement, we decreased the circulation gas pressure down to
$-0.071$~MPa, at first. Then, we added Xe gas~(with some amount of Rn) into the system to increase the inner gas pressure.
Figure~\ref{fig_press_dep} shows the result of these measurements.
In this estimation, only the data points in which Rn concentration is
more than $80~\mathrm{Bq/m^{3}}$ was selected to reduce the Rn
concentration dependence of the Rn adsorption efficiency.

We found an apparent increase in the Rn adsorption efficiency when the
circulation gas pressure is low.

\subsection{Comparison among ACF types} \label{sec_result}

We have compared the Rn adsorption efficiency among different ACF types.
In this comparison, the measurement data were selected as follows.
The circulation gas pressure is atmospheric pressure for all the data to
eliminate pressure dependence of the Rn adsorption efficiency.
The flow rate is between $0.14$~SLM and $1.4$~SLM, in which
no dependence on the Rn adsorption efficiency was observed.
In this comparison, we normalized the Rn adsorption efficiency values
by weight, to the typical weight of ACF sample in the main trap ($=12$~g),
since the amount of the ACF samples are different.
We could not adjust the Rn concentration level in each measurement,
though we observed the Rn concentration dependence in the region below 
$50~\mathrm{Bq/m^{3}}$. 
Therefore, this comparison was done as a function of the Rn concentration. 

\begin{figure}[tbh]
\centering\includegraphics[width=5in]{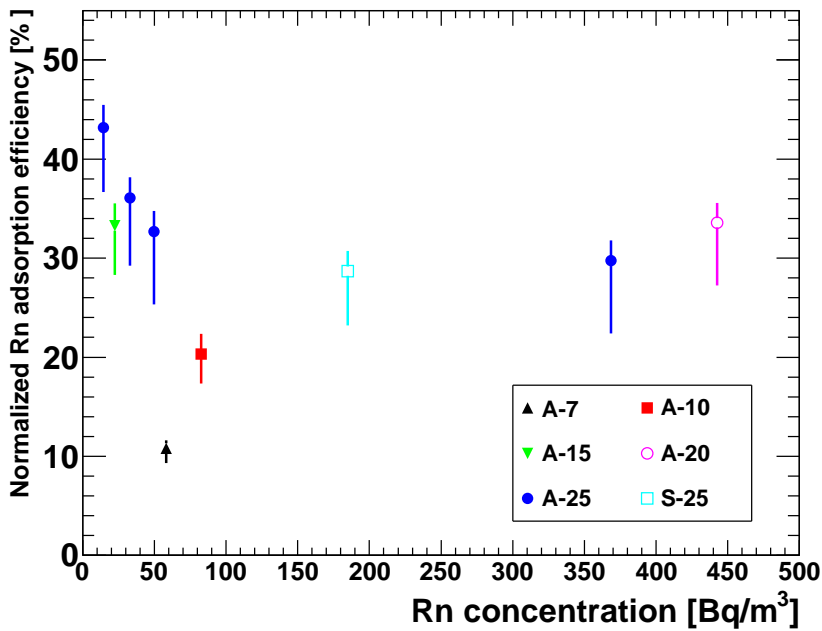}
 \caption{
 Normalized Rn adsorption efficiency  as a function of Rn
 concentration just before the adsorption phase.
 The circulation gas pressure is atmospheric pressure, and the flow rate
 is between $0.14$~SLM and $1.4$~SLM.
  The temperature setting of the main refrigerator during the adsorption
 phase is fixed at $-95 \mathrm{^{\circ}C}$.
}
\label{fig_comp1}
\end{figure}

\begin{figure}[tbh]
\centering\includegraphics[width=5in]{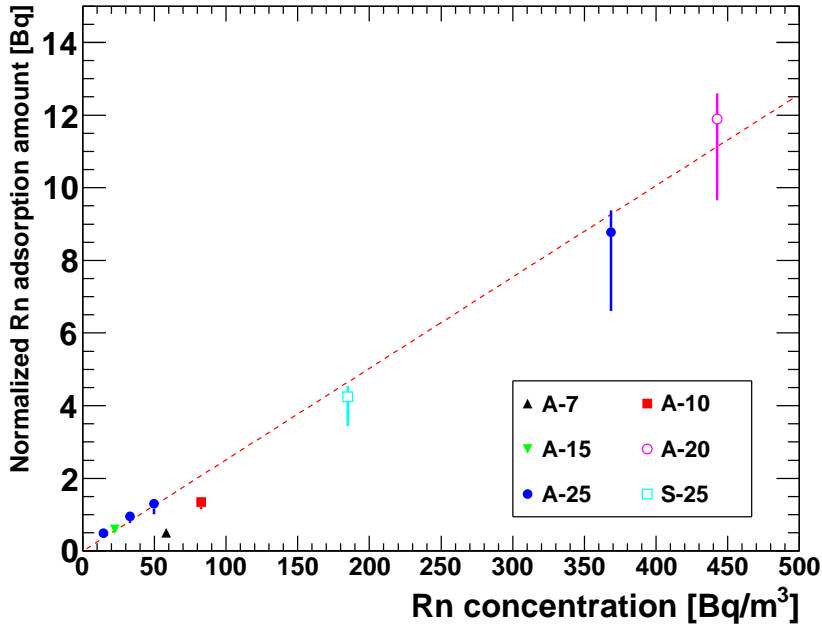}
 \caption{
 Normalized Rn adsorption amount as a function of Rn
 concentration just before the adsorption phase.
 The red dashed line is a fitted line of A-25 data.
 The used measurements are same as Fig.~\ref{fig_comp1}
 }
\label{fig_comp2}
\end{figure}

Figure~\ref{fig_comp1} shows a summary of this comparison.
In  Figure~\ref{fig_comp1}, several A-25 measurements are shown as a reference
in different Rn concentration regions. 
Table~\ref{table_result} summarizes numerical information for typical data
points shown in Figure~\ref{fig_comp1}.
\begin{table}[tb]
 \caption{Summary of the numerical information of typical measurements.
 The columns from the left side correspond to
 ACF type,
 amount of used ACF sample in the main trap,
 Rn concentration just before the adsorption phase,
 pressure drop during the adsorption phase,
 observed Rn adsorption efficiency $R$,
 and weight-normalized Rn adsorption efficiency, respectively.
 The errors are statistical only if it is not specified.}
\label{table_result}
\centering
\begin{tabular}{c||c|c|c|c|c}
\hline
Type & Amount & Rn     & P. drop & $R$ & Normalized efficiency \\ 
   & [g]  & [Bq/m$^3$] & [MPa] & [\%] & [\% ] \\ 
\hline \hline
A-7  & 29.97 & $58.1 \pm 0.9$ & $0.003$ & $27.0 \pm 0.4$ & $10.8 ^{+0.8} _{-1.5}$(stat.+syst.) \\ \hline
A-10 & 12.22 & $82.5 \pm 0.5$ & $0.002$ & $20.7 \pm 0.5$ & $20.3 ^{+2.1} _{-3.0}$(stat.+syst.) \\ \hline
A-15 & 11.11 & $22.4 \pm 0.5$ & $0.004$ & $30.8 \pm 0.6$ & $33.3 ^{+2.3} _{-5.0}$(stat.+syst.) \\ \hline
A-20 & 12.01 & $442.8\pm 0.8$ & $0.006$ & $33.6 \pm 0.2$ & $33.6 ^{+2.1} _{-6.4}$(stat.+syst.) \\ \hline
A-25 & 11.90 & $368.7\pm 1.2$ & $0.007$ & $29.5 \pm 0.3$ & $29.7 ^{+2.1} _{-7.4}$(stat.+syst.) \\ \hline
A-25 & 11.90 & $14.3 \pm 0.2$ & $0.006$ & $42.8 \pm 1.1$ & $43.2 ^{+2.4} _{-6.5}$(stat.+syst.) \\ \hline
S-25 & 11.88 & $184.0\pm 0.6$ & $0.005$ & $28.4 \pm 0.3$ & $28.7 ^{+2.1} _{-5.5}$(stat.+syst.) \\ \hline
\end{tabular}
\end{table}
In  Figure~\ref{fig_comp2}, the same measuremts are shown using normalized Rn adsorption amount. 
In these comparison plots, the dominant uncertainty is the pressure drop effect in the lower error bars.
Since it comes from the interpretation of the pressure drop phenomenon,
the uncertainties are taken into account among all data points.
Therefore the relative uncertainties among data points would be about
twice larger than the upper error bars. 
 
Then, comparing each ACF type to A-25 in a similar Rn concentration area, 
the normalized Rn adsorption efficiency (and amount) values of A-7, and A-10
are lower than that of A-25.
On the other hand, A-15, A-20, and S-25 show similar efficiency (and
amount) values as that of A-25.
Among these measurements, A-20 shows a slightly higher efficiency (and
amount) value
than A-25, though they are consistent within uncertainty.
The observed efficiency value of A-20 was $(33.6 ^{+2.1} _{-6.4}$) \%
(12 g of A-20, in atmospheric pressure Xe and $442.8 \pm 0.8$ Bq/m$^3$ Rn).

Among A-7, A-10, A-15, and A-20, the adsorption performance
looks increasing in this order.
From Table~\ref{table_acf}, specific surface area, pore volume,
and meso pore volume ratio are also increasing in the same order.
Since kinetic diameters~(Lennard-Jones parameter $\sigma$) of Air, Ar, Xe, and Rn gas molecules are
estimated as $0.352$--$0.369$~nm, $0.340$--$0.346$~nm, $0.392$--$0.410$~nm,
and $0.417$--$0.421$~nm, respectively~\cite{molecule1, molecule2, molecule3},
and typical range of van-der-Waals interaction is about $1$~nm,
the relevant pore size on the adsorption would be around $1$--$2$~nm.
Therefore, the meso pores~($2$--$50$~nm) would not affect directly
on the adsorption process while the increase of specific surface area and pore volume
would be effective in this comparison.

About A-25 and S-25, the specific surface area and pore volume
are larger than A-20, but the measured adsorption performance 
looks not improved.
A possible reason would be the average pore diameter
are too large to adsorb Xe atom, since the intensity of
van-der-Waals force is inversely proportional to the cube of distance.

\section{ Conclusion }

We have carried out Rn adsorption measurements with ACFs using a newly
developed test bench at Kobe University, Japan.
The radioactivity of the ACFs provided from UNITIKA Ltd. was lower or
comparable to that of the activated charcoals used in previous researches. 
The measured radioactivity of A-20 for the uranium series
was $<5.5$~$\mathrm{mBq/kg}$ with $90\%$ confidence level.
In Air and Ar gas, ACF A-15 demonstrated an
excellent adsorption efficiency of $1/10000$ reduction at maximum before the saturation of Rn adsorption,
 and more than $97\%$ adsorption efficiency after the saturation.
In Xe gas, the adsorption efficiency values are lower than that in Air
or Ar gas. 
We carried out a set of measurements on Rn adsorption in Xe gas under
various conditions, then estimated ACF's basic performance of flow rate,
concentration, and pressure dependence of the Rn adsorption ability in
Xe gas. 
We also observed some difference on Rn adsorption ability from different
ACF types. Among the tested ACFs, A-20 has the best Rn
adsorption ability.
The observed Rn adsorption efficiency was $(33.6 ^{+2.1} _{-6.4}$) \%
(12 g of A-20, in atmospheric pressure Xe and $442.8 \pm 0.8$ Bq/m$^3$ Rn).
S-25, A-25, and A-15 also show similar adsorption performance.

\section*{Acknowledgment}
The authors would greatly appreciate Prof. K. Kaneko at
Research Initiative for Supra-Materials (RISM), Shinshu University
for his very helpful suggestions on this research topic.
The authors also would greatly appreciate UNITIKA Ltd. for their kind
support on this research.
They provided various ACF samples, including some special test products.
This work is supported by JSPS
KAKENHI Grand Number JP26104008, 18H05536, 16H03973, and the joint research
program of the Institute for Cosmic Ray Research~(ICRR), the University
of Tokyo.


\end{document}